\documentclass[%
APS,
twocolumn,
reprint,
superscriptaddress,
 amsmath,amssymb,
 aps,
]{revtex4}

\usepackage{amssymb,amsmath}
\usepackage{epsfig}
\usepackage{graphics}
\usepackage{graphicx}
\usepackage{dcolumn}
\usepackage{bm}
\usepackage{epstopdf}
\usepackage{color}
\usepackage{lineno}
\modulolinenumbers[5]

\begin{document}
	
\title{Enhanced photogalvanic effect in the two-dimensional MgCl$_2$/ZnBr$_2$ vertical heterojunction by inhomogenous tensile stress}

\author{Liyu Qian}
\affiliation{Department of Physics, Shanghai Normal University, 100 Guilin Road, Shanghai 200234, P. R. China}
\author{Juan Zhao}
\affiliation{Department of Physics, Shanghai Normal University, 100 Guilin Road, Shanghai 200234, P. R. China}
\author{Yiqun Xie}
\email{yqxie@shnu.edu.cn}
\affiliation{Department of Physics, Shanghai Normal University, 100 Guilin Road, Shanghai 200234, P. R. China}

\begin{abstract}
 The photogalvanic effect (PGE) occurring in noncentrosymmetric materials enables the generation of a dc photocurrent at zero bias with a high polarization sensitivity, which makes it very attractive in photodetection. However, the magnitude of the PGE photocurrent is usually small, leading to a low photoresponsivity, and therefore hampers its practical application in photodetection. Here, we propose an approach to largely enhancing the PGE photocurrent by applying an inhomogenous mechanical stretch, based on quantum transport simulations. We model a two-dimensional photodetector consisting of the wide-bandgap MgCl$_2$/ZnBr$_2$ vertical van der Waals heterojunction with the noncentrosymmetric $C_{3v}$ symmetry. Polarization-sensitive PGE photocurrent is generated under the vertical illumination of linearly polarized light. By applying inhomogenous mechanical stretch on the lattice, the photocurrent can be largely increased by up to 3 orders of magnitude due to the significantly increased device asymmetry. Our results propose an effective way to enhance the PGE by inhomogenous mechanical strain, showing the potential of the MgCl$_2$/ZnBr$_2$ vertical heterojunction in the low-power UV photodetection.
\end{abstract}

\maketitle
\section{\textbf{INTRODUCTION}}
Ultraviolet (UV) photodetectors (PDs) have been widely utilized in military and civilian fields, such as missile detection, environmental monitoring, and space optical communications \cite{AM2016-H.Chen,AP2019-Q.Zhang,ACSNano2018-D.Guo}. Robust photoresponsivity, high polarization sensitivity, low-energy power consumption, ultra-small size and good flexibility are important figure of merits for the UV photodetections. In recent years, UV-PDs based on two-dimensional (2D) semiconductors have shown great potential to fulfill these demands. There has been considerable research on UV-PDs based on the 2D wide bandgap (2D-WBG) semiconductors, such as MgO with an energy gap (Eg) of 7.3 eV \cite{ACSAMI2018-W.Zhang-MgO} few-layered h-BN \cite{ACSAMI2018-W.Zhang-BN} and FePS$_3$ \cite{Nanotechnology2018-Y.Guo}. Outstanding performances have been achieved via such 2D-WBG UV-PDs. For example, the graphene/ZnS/4H-SiC photodetector has demonstrated an ultrafast photoresponse to 250 nm light with a 28 ms rise time \cite{ACSAMI2019-H.Kan}.

Various mechanisms are responsible for the generation of the photocurrent in these photodetectors, which include the photovoltaic effect, photogating, photoconducting, as well as photothermal effects. Of particular importance is the photogalvanic effect (PGE) that occurs in materials without the space inversion symmetry under the illumination of polarized light \cite{Xiao2020,PRB1981-R.von,PRA1978-V.I.Belinicher,JPCM2003-S.D.Ganichev,PRB2020-X.Tao,Nanoscale2020-M.M.R.Moayed,AcsNano2020-M.Chen,SciAdv2020-C.Guo}. The PGE is able to generate a dc photocurrent that is highly polarization sensitive, without the need for any external electric fields or a pn-junction. Therefore, the PGE is a perfect mechanism for the self-powered and polarization sensitive UV photodetection with a low dark current. In our previous work, we have proposed the UV photodetection based on the PGE by using the photodetector consisting of the wide-bandgap 2D materials and their heterostructures \cite{JMCA2019-Y.Luo}. A recent experiment has also demonstrated that the PGE is an effective mechanism for the UV photodetection with a high polarization sensitivity and low dark current \cite{ACIE2020-Y.Peng}.

However, the magnitude of the PGE photocurrent is usually small, which leads to a relatively low photoresponsivity typically of mA/W. Therefore, it is of practical importance to find a mechanism to enhance the magnitude of the PGE photocurrent. Several recent experiments have suggested that the PGE photocurrent can be largely enhanced under mechanical strains \cite{NATURE2019-Y.J.Zhang,NatPhoto2016-J.E.Spanier,APL2016-J.Li}. It has been shown that, by applying an appropriate mechanical stress gradient on the SrTiO$_3$ single crystal, the PGE photocurrent can be enhanced by orders of magnitude \cite{Science2018-M.-M.Yang}, which is known as the flexophotovoltaic efect, though the controversy was raised regarding on the underlying mechanism \cite{ACSnano2019-H.Zou}. Direct experimental evidence for the efective enhancement of the PGE by mechanical strain has recently been achieved in the Fe-doped LiNbO$_3$ single crystal, in which the photocurrent was increased by 75$\%$ even under a tiny uniaxial compressive stress (0.005$\%$) on the lattice \cite{Sciadv2010-S.Nadupalli}. Motivated by these experimental findings, we have investigated the influence of uniform mechanical tensile stress on the PGE for a phosphonene photodetector by using quantum transport simulations, and proposed that the large enhancement of the PGE photocurrent is mainly due to the strain-increased device asymmetry \cite{ZHAOJUAN-PRA}.

In this work, using quantum transport simulations, we present a theoretical study on the PGE in the photodetector based on the 2D MgCl$_2$/ZnBr$_2$ vertical van der Waals heterojunction (vdWH). By applying an inhomogenous mechanical tensile stress on the photodetector, the PGE photocurrent can be largely enhanced under the illumination of linearly polarized light. Our results propose an effective approach to enhancing the PGE, suggesting its promising application in the low-power UV photodetection.
{\normalsize }
\section{Model and Methods}
The 2D monolayer MgCl$_2$ and ZnBr$_2$ have recently been predicted with a stable configuration \cite{NatNanotechnol2018-N.Mounet}. Their primitive cells contain two haloid atoms (Cl, Br) and a metal atom (Mg, Zn), as shown in Figs.1(a) and 1(b). The lattice constant is 3.645 {\AA} and 3.761 {\AA} for MgCl$_2$ and ZnBr$_2$, with the bandgap of 6.0 eV and 3.41 eV, respectively, according to our previous first-principle calculations \cite{JMCA2019-Y.Luo}. Since the monolayer MgCl$_2$ and ZnBr$_2$ have a centrosymmetric $D_{3d}$ symmetry, the PGE cannot be generated. Therefore, to obtain the PGE, we construct a vdWH using the monolayer MgCl$_2$ and ZnBr$_2$, as shown in Fig.1. This vdWH has the noncentrosymmetric $C_{3v}$ symmetry, and therefore the PGE can be generated under the vertical illumination of linearly polarized light.

The electronic bandstructure of the MgCl$_2$/ZnBr$_2$ vdWH are calculated using the Vienna Ab-initio Simulation Package (VASP) \cite{PRB1996-G.Kresse,PRB1999-G.Kresse}. The plane wave basis set with the projector augmented wave (PAW) pseudopotential is adopted. The exchange correlation potentials are approximated using the generalized gradient approximation (GGA) functional as parameterized by Perdew, Burke and Ernzerhof (PBE) functional \cite{PRL1996-J.P.Perdew}. Brillouin zone integration is performed with a 15$\times$1$\times$15 k-point mesh for geometry optimization and self-consistent electronic structure calculations. The kinetic energy cutoff for the plane wave basis is 500 eV. All the atoms in the unit cell are fully relaxed until the force on each atom is less than 0.001 eV/\AA\, and the convergence criteria for energy in the self-consistent field cycle was $10^{-6}$ eV. The supercell is periodic in the $xz$-plane and is separated by a 20 {\AA} vacuum layer in the $z$-direction to avoid the interaction between the imaginary cells.

The photocurrent is calculated using the first-principles quantum transport package NanoDcal \cite{NANODCAL}, where density functional theory is used combined with nonequilibrium Green's formalism (NEGF-DFT). The photocurrent is calculated based on linear response theory \cite{JAP2002-L.E.Henrickson}. Specifically, for linearly polarized light, the photocurrent injecting into the left lead (electrode) can be written as {\tiny }\cite{PRA2018-Y.Xie,Nanotechnology2015-Y.Xie,JMCA2019-Y.Luo},
\begin{eqnarray}\label{LPGE1}
J_L^{(ph)}&=&{ie\over h}\int\{\cos^2\theta\textrm{Tr}\{\Gamma_L[G^{<(ph)}_{1}+f_L(G^{>(ph)}_{1}-G^{<(ph)}_{1})]\}  \nonumber\\
&+&\sin^2\theta\textrm{Tr}\{\Gamma_L[G^{<(ph)}_{2}+f_L(G^{>(ph)}_{2}-G^{<(ph)}_{2})]\} \nonumber\\
&+&{2\sin(2\theta)}\textrm{Tr}\{\Gamma_L[G^{<(ph)}_{3} \nonumber\\
&+&f_L(G^{>(ph)}_{3}-G^{<(ph)}_{3})]\}\}\textrm{d}E, \nonumber
\end{eqnarray}

where $G^{>/<(ph)}_{1,2,3}$ denotes the greater/lesser Green's functions with electron-photon interaction, which are determined by the symmetry, photon frequence and polarization vector $\textbf{e}$. For linearly polarized light, the polarization vector \textbf{e}=$\cos\theta$\textbf{e}$_1$+$\sin\theta$\textbf{e}$_2$, where $\theta$ is the angle formed by the polarization direction with respect to the vector $\textbf{e}_1$. The vectors $\textbf{e}_1$ and $\textbf{e}_2$ are set along the zigzag and armchair directions, respectively. The photocurrent can be normalized as, $I_{ph}= J_L^{(ph)}/eI_\omega$ which still has dimensions of area of $a_0^2$/photon, where $a_0$ is the Bohr radius. In the numerical calculations, the DZP atomic orbital basis is used to expand all the physical quantities; the exchange and correlation were treated at the level of the GGA functional as parameterized by the PBE approximation; atomic cores are determined by the standard norm conserving nonlocal pseudopotentials, and 16$\times$1$\times$1  k-points are used. These calculation details are verified to obtain converged results.

According to our previous work \cite{ZHAOJUAN-PRA}, the magnitude of the PGE photocurrent is determined by the $|J_L^{(ph)}|=\Delta J=|J_{+} -|J_{-}||$, where $J_{-}= {ie\over h}\int^0_{-\infty} \textrm{Tr}[\Gamma_L G^{>(ph)}]\textrm{d}E <0$ denotes  the current flowing out the center region to  the left lead, while $J_{+}= {ie\over h}\int_0^{+\infty} \textrm{Tr}[\Gamma_L G^{<(ph)}]$ $ \textrm{d}E$ $>0$ denotes the current flowing into the center region from the lead.  Therefore we can define a factor $A$ to describe the device asymmetry as
\begin{equation}
\textit{A} =  \left| \frac{J_{+}-|J_{-}|}{J_{+}} \right|. \label{I}
\end{equation}
For the device with the space inversion symmetry the current flowing into  the center region equals that out of the center region, that is, there is no net current flowing through the device. In this case,  $J_{+}=|J_{-}|$, so that $J_L^{(ph)}=0$, and  $A$=0. For the device without the inversion symmetry, there is a net photocurrent and  $A$ is nonzero. The magnitude of the photocurrent can then be rewritten as,
\begin{equation}
|J_L^{(ph)}|= AJ_{+}. \label{J}
\end{equation}
In the below, we will show that the device asymmetry plays a dominant role in enhancing the photocurrent under inhomogenous mechanical stress.

\section{Results and discussion}
We consider three configurations of the 2D MgCl$_2$/ZnBr$_2$ vertical vdWHs, namely, AA, AB and AC, as shown in Fig.1. The lattice constant of the ZnBr$_2$ (3.761 {\AA}) is used for the heterostructure, which means a 3.18$\%$ stretch in the lattice of the MgCl$_2$. For the AA structure, the Zn atoms in the top layer of ZnBr$_2$ lie directly above the Mg atoms in the bottom layer of MgCl$_2$, as shown in Fig.1(a). For the AB structure, the Zn atoms of the top ZnBr$_2$ layer lie directly above the Br atoms in the top-sublayer of the MgCl$_2$ (Fig.1(b)). For the AC structure, Zn atoms of the top ZnBr$_2$ layer are directly above the Br atoms of the bottom MgCl$_2$ sublayer (Fig.1(c)). We calculate the binding energy of the three heterojunctions, which is defined as $E_B$=$E_1+E_2$-$E_{vdWH}$. Here, $E_{vdWH}$, $E_1$ and $E_2$ represent the total energy of the heterojunction, monoalyer MgCl$_2$ and ZnBr$_2$, respectively. Considering the negative signs of the energies, a higher (more positive) binding energy means a more strongly bounded and stable system. The AA structure is the most stable one among the three vdWHs, as indicated in Tab.1. Therefore, in the following we mainly investigate electronic properties and the PGE of the AA MgCl$_2$/ZnBr$_2$ vdWH.
\begin{figure}[hptb]
	\begin{center}
		\includegraphics[scale=0.28]{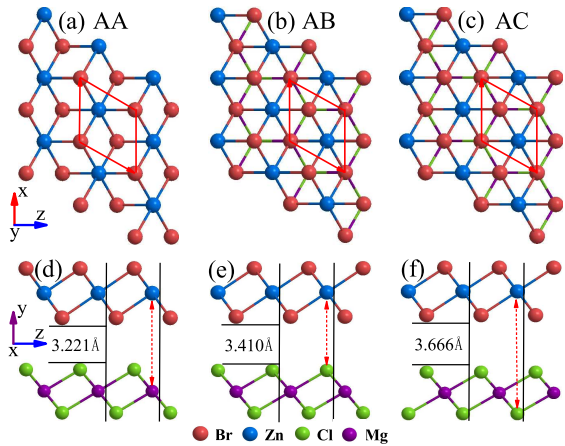}
	\end{center}
	\caption{Three kinds of stacking patterns of the 2D MgCl$_2$/ZnBr$_2$ vertical vdWH. (a), (b) and (c) are top views for AA, AB and AC configurations, respectively, and (d), (e) and (f) are corresponding side views. The red arrows in (a) denote the lattice constants of the primitive cell. }
	\label{fig1}
\end{figure}

\begin{table}[ht]
	\small
	\caption{Total energy ($E_{vdWH}$), binding energy ($E_B$), interlayer spacing $d $ and lattice constant $a$ for three kinds of 2D MgCl$_2$/ZnBr$_2$ vertical vdWHs. }
	\label{tbl:example2}
		\begin{tabular*}{0.48\textwidth}{@{\extracolsep{\fill}}lllll}
			\hline
			Stacking &$E_{vdWH}$(eV) &$E_B$(eV) & $d$({\AA}) & $a$({\AA})  \\
			\hline
			AA & -5.976 & 0.222 & 3.221 & 3.761 \\
			AB & -5.969 & 0.214 & 3.410 & 3.761 \\
			AC & -5.920 & 0.166 & 3.666 & 3.761 \\
			\hline
		\end{tabular*}\label{tab1}
\end{table}

The electronic bandstructure of the AA MgCl$_2$/ZnBr$_2$ vertical vdWH is shown in Fig.2(a). It has an indirect bandgap of 3.47 eV, with the valence band maximum (VBM) located at Gamma point, while the conduction band minimum (CBM) is located at M point. Figure 2(b) shows the absorption coefficient $\alpha(\omega)$ of the AA MgCl$_2$/ZnBr$_2$ vdWH. There are two prominent absorption peaks located in the ultraviolet range, and the first peak appears at around 5.1 eV. We will focus on the PGE for the photon energy lower than 6.0 eV, which covers the first absorption peak.
\begin{figure}[hptb]
	\begin{center}
		\includegraphics[scale=0.30]{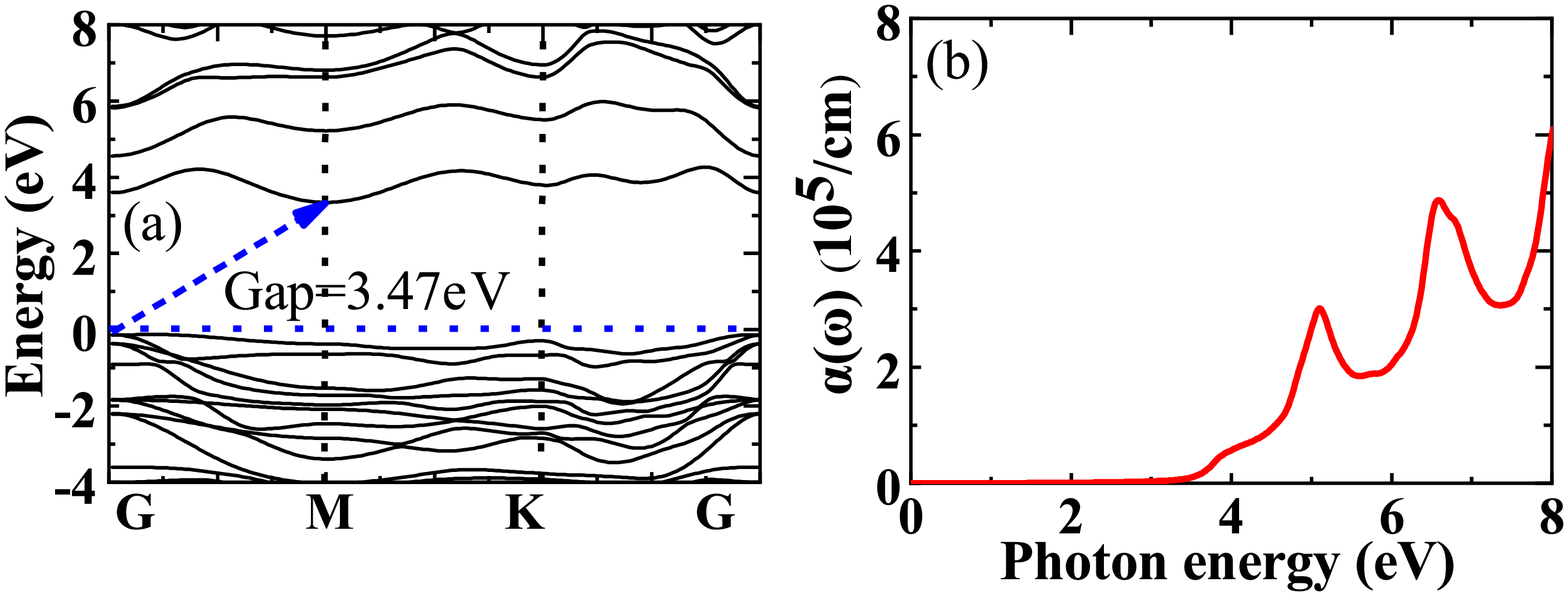}
	\end{center}
	\caption{(a) The electronic bandstructure and (b) optical absorption coefficient of the AA MgCl$_2$/ZnBr$_2$ vdWH. }
	\label{fig2}
\end{figure}

We model an UV-PD by using the AA MgCl$_2$/ZnBr$_2$ vdWH, which contains: left and right electrodes (leads), and the central region, as indicated in Fig.3(a). The whole device is periodical in the $x$-direction, and two electrodes are extended into $z=\pm\infty$, respectively. When the center region is illuminated by linearly polarized light, the PGE is generated due to the $C_{3v}$ symmetry, giving rise to a robust photocurrent that flows along the $z$ direction. The photon energy considered is from 3.5 to 6.0 eV, which is greater than the bandgap of the vdWH and covers the first optical absorption peak.
\begin{figure}[hptb]
	\begin{center}
		\includegraphics[scale=0.28]{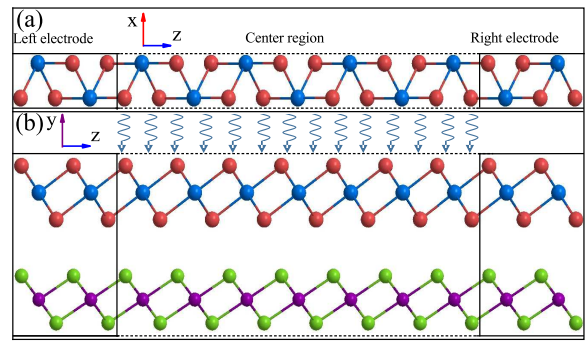}
	\end{center}
	\caption{Atomic model for the photodetector composed of the AA MgCl$_2$/ZnBr$_2$ vertical vdWH. (a) and (b) are top and side views of the photodetector, respectively. The blue curved arrows in (b) represent the vertically incident linearly polarized light on the center region. }
	\label{fig3}
\end{figure}

\begin{figure}[hptb]
	\begin{center}
		\includegraphics[scale=0.30]{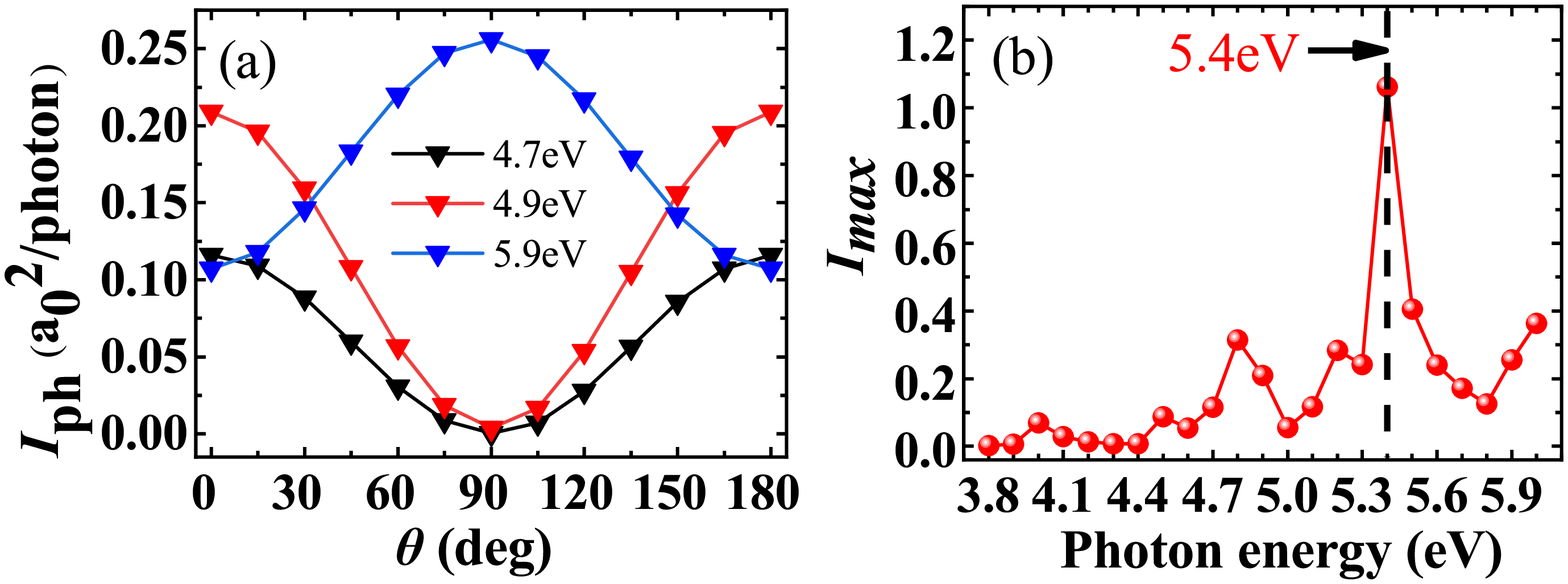}
	\end{center}
	\caption{(a) Variation of the photocurrent with the polarization \label{key}angle $\theta$ for photon energies of 4.7, 4.9 and 5.9 eV. (b) The maximum photocurrent for different photon energies. }
	\label{fig4}
\end{figure}

Our calculations show that the PGE photocurrent has a form $\beta*\cos2\theta+I_0$, which is determined by the $C_{3v}$ symmetry, in agreement with the phenomenological theory of the PGE \cite{SovPhysUsp1980-V.I.Belinicher,APL2000-S.D.Ganichev}. Here, $\beta$ and $I_0$ are determined by the symmetry and photon energy. Fig.~\ref{fig4}(a) gives the photocurrent for photon energies of 4.7 eV, 4.9 and 5.9 eV as examples, which shows an evident cosine dependence on the polarization angle $\theta$. We can then obtain the maximum photocurrent ($I_{max}$) at either $\theta=0^{\circ}$ or $\theta=90^{\circ}$ for each photon energy. The $I_{max}$ for different photon energies is shown in Fig.~\ref{fig4}(b). For photon energies of 3.5, 3.6, and 3.7 eV, the photocurrent is very small with a magnitude of $10^{-4}$ (not shown), and from 3.8 eV the photocurrent begins to increase. The photocurrent reaches the maximum (1.06) at 5.4 eV. Overall, the photocurrent in the higher photon energy range ($>$ 4.5 eV) is larger than that in the low photon energy range (3.5-4.5 eV), which should be partly attributed to a stronger optical absorption for high photon energies, as shown in Fig.2(b).

Next, we investigate the influence of mechanical tension stress on the PGE photocurrent. Three types of tensile stress are exerted on the lattice. The first one is a uniform 5$\%$ stretch on the entire device along the z-direction. The second one is a partial stretch, in which only the right part of the device is stretched by 5$\%$ along the z-direction. The third is a gradient stretch: The left electrode and the left 2/7 of the center region are invariant; the following 3/7 of the center region is stretched by 3$\%$ along the z-direction; the rest 2/7 of the center region and the right electrode are stretched by 5$\%$. For the uniform stretch, the $C_{3v}$ symmetry of the photodetector is not changed, so that the photocurrent remains the cosine dependence on the 2$\theta$, as shown in Fig.5(a). For the partial and gradient stretch, the symmetry of the photodetector is reduced from the $C_{3v}$ to $C_{s}$, while the photocurrent still holds the cosine form under the $C_{s}$ symmetry, as shown in Fig.5(b), which is in a good agreement with the phenomenological theory of the PGE. The maximum photocurrent for different photon energies is shown in Figs.5(c). It can be seen that the photocurrent under partial and gradient stretches (the blue and red spheres) is much larger than that of the uniform stretch. Here, we calculate the increase ratio $R_I$ of the photocurrent, which is the ratio of the photocurrent under mechanical strain with respect to that without any strain, as shown in Figs.5(d) The maximum $R_I$ is 1372 at 3.8 eV for the gradient stretch, and is 849 at 3.8 eV for the partial stretch. For a number of photon energies, e.g., 4.1 to 4.4 eV, $R_I$ is greater than 100, and moreover, for all photon energies, $R_I$ is greater than 1 for both the partial and gradient stretches. In contrast, for uniform stretch the photocurrent only increases for a few of low photon energies, and is even decreased for some high energies. These results show that the magnitude of the photocurrent can be largely enhanced by partial and gradient stretches. The reason is that the device is reduced from the $C_{3v}$ symmetry to the $C_{s}$ symmetry under partial and gradient stretches, which means that the device becomes more asymmetric, in other words, the device asymmetry is increased. The magnitude of the photocurrent has a linear dependence on the device asymmetry, as proposed in our previous work \cite{ZHAOJUAN-PRA}, and therefore, the photocurrent is largely enhanced. In contrast, for the uniform stretch the symmetry of the photodetector is not changed, so that the magnitude of the photocurrent does not change considerably.
\begin{figure}[hptb]
	\begin{center}
		\includegraphics[scale=0.42]{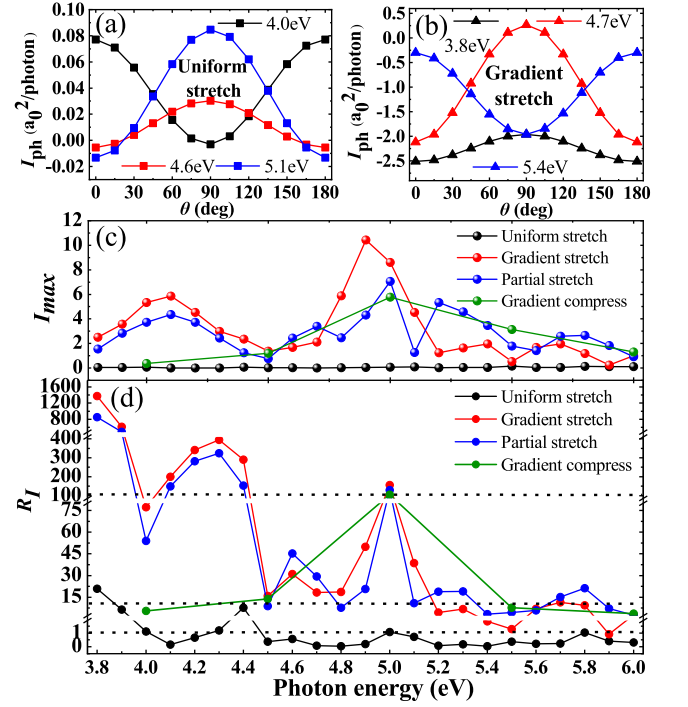}
	\end{center}
	\caption{Variation of the photocurrent with the polarization angle  $\theta$ for (a) the uniform stretch and (b) gradient stretch. (c) The maximum photocurrent at different photon energies for three types of stretches. (d) The increase ratio $R_I$ of the photocurrent under stretch with respect to that without the stretch.  }
	\label{fig5}
\end{figure}

\begin{figure}[hptb]
	\begin{center}
		\includegraphics[width=8cm]{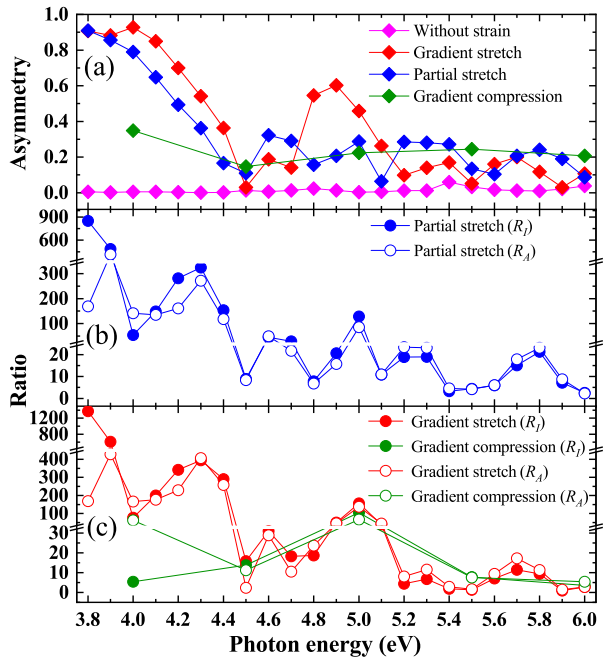}
	\end{center}
	\caption{(a) The device symmetry $A$ at different photon energies for different mechanical strain. (b) Comparison of the $R_I$ with the increase ratio of the device asymmetry ($R_A$) for   partial stretch, and (c) for gradient stretch and compression.  }
	\label{fig6}
\end{figure}

Then, we calculate the device asymmetry $A$, and also the increase ratio $R_A$ of the device asymmetry with respect to that without any strain. Fig.~\ref{fig6}(a) plots the device symmetry $A$ at different photon energies for different mechanical strain, which shows that the device asymmetry is considerably increased, as compared to that without any strain (the pink squares).
Furthermore, we compare $R_I$ with $R_A$  in Figs.~\ref{fig6}(b) and 6(c) at different photon energies for different mechanical strain. It can be seen that the overall trend of the $R_I$ agrees very well with that of the $R_A$ for almost all of the photon energies. We also examined the photocurrent under gradient compression for several photon energies, and found that the photocurrent is also significantly enhanced, with a largest $R_I$ of 128 at 5.0 eV, as shown in Figs.~\ref{fig5}(c) and 5(d)(see the green circles).
This is also attributed to the largely increased device asymmetry (see Figs.~~\ref{fig6}(a) and 6(c)). These results strengthen that the inhomogenous strain including partial stretch, gradient stretch and gradient compression can effectively enhance the PGE photocurrent in the photodetector because of the reduced device symmetry.

\section{CONCLUSIONS}
In conclusion, we have studied the PGE in the photodetector composed of the 2D MgCl$_2$/ZnBr$_2$ vertical vdWH by using quantum transport simulations. PGE photocurrent is generated at zero bias under the illumination of linearly polarized light in the UV range for photon energies from 3.8 eV to 6.0 eV, owing to the noncentrosymmetric $C_{3v}$ symmetry. By applying partial stretch (5$\%$) and gradient stretch along the transport direction on the device, the PGE photocurrent is largely enhanced by up to 3 orders of magnitude for all of the photon energies. This is because the partial and gradient stretches reduce the device symmetry from the $C_{3v}$ to the $C_{s}$, which increases the device asymmetry, and therefore the photocurrent. Our results propose an effective way to largely enhance the PGE by inhomogenous tensile stress, and suggest a potential application of the 2D MgCl$_2$/ZnBr$_2$ vertical vdWH in the low-power UV photodetection.

\section*{\textbf{ACKNOWLEDGMENT}}
This work is supported by the National Natural Science Foundation of China under Grant No.51871156.


\end{document}